\documentstyle[11pt,newpasp,psfig,epsfig,twoside]{article}
\begin{document}
\newcommand {\ignore}[1]{}
\def\tp{these proceedings}
\def\tts{talks at this session}
\def\ts{talk at this session}
%\def{\ss}{\scriptstyle}
%%%
\def\smallfrac#1#2{{\textstyle{#1 \over #2}}}
\def\lsim{\:\raisebox{-0.5ex}{$\stackrel{\textstyle<}{\sim}$}\:}
\def\gsim{\:\raisebox{-0.5ex}{$\stackrel{\textstyle>}{\sim}$}\:}
\def\VEV#1{\left\langle #1\right\rangle}
\def\fig#1{{Fig. (\ref{#1})}}
\def\smallfrac#1#2{{\textstyle{#1 \over #2}}}
\def\thefootnote{\fnsymbol{footnote}}
\def\N{$\cal N$ }
\def\slash#1{#1\!\!\! /}
\def\nn{\nonumber}
\def\rp{$R_p \hspace{-1em}/\;\:$ }
\def\Eq#1{{Eq. (\ref{#1})}}
\def\eq#1{{eq. (\ref{#1})}}
\def\Fig#1{{Fig. (\ref{#1})}}
\def\be{\begin{equation}}
\def\ee{\end{equation}}    
\def\bear{\be\begin{array}}
\def\eear{\end{array}\ee}
\def\bea{\begin{eqnarray}}
\def\eea{\end{eqnarray}}
\def\SM{Standard Model }
\def\baselinestretch{1.3}
\def\vb#1{\vbox to #1 pt{}}
\def\beqa{\begin{eqnarray}}
\def\eeqa{\end{eqnarray}}
\def\ni{\noindent}
\def\ba{\begin{array}}
\def\ea{\end{array}}
\def\ovl{\overline}
\def\ds{\displaystyle}
\def\epjc#1#2#3{{\it Eur.\ Phys.\ J. }{\bf C #1} (#2) #3}
\def\npb#1#2#3{{\it Nucl.\ Phys.\ }{\bf B #1} (#2) #3}
\def\plb#1#2#3{{\it Phys.\ Lett.\ }{\bf B #1} (#2) #3}  
\def\prd#1#2#3{{\it Phys.\ Rev.\ }{\bf D #1} (#2) #3}
\def\prep#1#2#3{{\it Phys.\ Rep.\ }{\bf #1} (#2) #3}
\def\prl#1#2#3{{\it Phys.\ Rev.\ Lett.\ }{\bf #1} (#2) #3}
\def\mpla#1#2#3{{\it Mod.\ Phys.\ Lett.\ }{\bf A #1} (#2) #3}
\def\sjnp#1#2#3{{\it Sov.\ J.\ Nucl.\ Phys.\ }{\bf #1} (#2) #3}
\def\jetpl#1#2#3{{\it Sov.\ Phys.\ JETP Lett.\ }{\bf #1} (#2) #3}
\def\rnc#1#2#3{{\it Riv. Nuovo Cimento }{\bf #1} (#2) #3}
\def\yf#1#2#3{{\it Yad.\ Fiz.\ }{\bf #1} (#2) #3}
\def\hepph#1{{\tt hep-ph/#1}}
%%%
\def\e6{$E(6)$} 
\def\10{$SO(10)$}
\def\21{$SU(2) \otimes U(1) $}
\def\lr{$SU(2)_L \otimes SU(2)_R \otimes U(1)$}
\def\422{$SU(4) \otimes SU(2) \otimes SU(2)$} 
\def\321{$SU(3) \otimes SU(2) \otimes U(1)$}
%%%
\def\neq{\hbox{$\nu_e$ }}
\def\nm{\hbox{$\nu_\mu$ }}
\def\nt{\hbox{$\nu_\tau$ }}        
\def\21{$SU(2) \otimes U(1)$}
\def\ie{{\it i.e.}}
\def\etal{{\it et al.}}
\def\half{{\textstyle{1 \over 2}}}
\def\third{{\textstyle{1 \over 3}}}
\def\quarter{{\textstyle{1 \over 4}}}
\def\sixth{{\textstyle{1 \over 6}}}
\def\eighth{{\textstyle{1 \over 8}}}
\def\sqrthalf{{\textstyle{1 \over \sqrt{2}}}}
\def\bold#1{\setbox0=\hbox{$#1$}
     \kern-.025em\copy0\kern-\wd0
     \kern.05em\copy0\kern-\wd0
     \kern-.025em\raise.0433em\box0 }
%%%%%%%%%%%%%%% End of My Macros %%%%%%%%%%%%%%%%%%%%%%%%%%%%%%%%%%%%   

 \newcommand{\wt}{\widetilde}
 \newcommand {\chiz} [1] {\tilde{\chi}^{0}_{#1} }
 \newcommand {\chiw} [1] {\tilde{\chi}^{\pm}_{#1} }

\def\rp{$R_p \hspace{-1em}/\;\:$ }
\def\Eq#1{{Eq. (\ref{#1})}}
\def\eq#1{{eq. (\ref{#1})}}
\def\beqa{\begin{eqnarray}}
\def\eeqa{\end{eqnarray}}

\def\vb#1{\vbox to #1 pt{}}

%%%%%%%%%%%%%%%%%%%%%%%%%%%%%%%%%%%%%%%%%%%%%

\markboth{J.~W.~F.~Valle}{Historical Development of Modern Cosmology }
\pagestyle{myheadings}
%\nofiles

% Some definitions I use in these instructions.

\def\emphasize#1{{\sl#1\/}}
\def\arg#1{{\it#1\/}}
\let\prog=\arg

\def\edcomment#1{\iffalse\marginpar{\raggedright\sl#1\/}\else\relax\fi}
\marginparwidth 1.25in
\marginparsep .125in
\marginparpush .25in
\reversemarginpar

\title{Status of Neutrino Oscillations}
 \author{J.~W.~F.~Valle}
\affil{Instituto de F\'{\i}sica Corpuscular -- C.S.I.C. -- Universitat de
 Val{\`e}ncia  \\
     Ed. de Institutos de Paterna -- Apartado de Correos 22085 -
    46071  Val{\`e}ncia, Spain}

\begin{abstract}
  Solar and atmospheric neutrino data require physics beyond the
  Standard Model of particle physics.
  The simplest, most generic, but not yet unique, interpretation of
  the data is in terms of neutrino oscillations.
  I summarize the results of the latest three-neutrino oscillation
  global fit of the data, in particular the bounds on the angle
  $\theta_{13}$ probed in reactor experiments.  Even though not
  implied by the data, bi-maximal neutrino mixing emerges as an
  attractive possibility either in hierarchical or
  quasi-degenerate neutrino scenarios.
\end{abstract}

\section{Introduction}

Undoubtedly the solar (Suzuki 2000) and atmospheric (Sobel 2000,
Becker-Szendy 1992) neutrino problems provide the two most important
milestones indicating physics beyond the Standard Model (SM).  Of
particular importance has been the confirmation in 1998 by the
Super-Kamiokande (SK, for short) collaboration of the
zenith-angle-dependent deficit of atmospheric neutrinos.  Altogether
the data provide a strong evidence for \neq and \nm conversions,
respectively.
Neutrino conversions are a natural consequence of theories beyond the
Standard Model (Valle 1991).  The first example is oscillations of
low-mass neutrinos.
While the theoretical understanding of the origin of neutrino masses
is still lacking, there is a variety of attractive options available.
Most likely, the exceptional nature of neutrinos as the only
electrically neutral fermions in the SM underlies the smallness of
their mass, as it would be associated with the violation of lepton
number.
Indeed in gauge theories one expects, on fundamental
grounds,  neutrinos to be Majorana fermions (Schechter 1980a). 
This is the generic situation in actual models.
It will be surprising indeed if massive neutrinos turn out to be Dirac
particles, like the quarks.
Lepton number violation would imply processes such as neutrino-less
double beta decay (Schechter 1982), novel CP violation effects
(Schechter 1980a and 1981a), and/or neutrino electromagnetic
properties (Schechter 1981b), so far unobserved.
Present data and theoretical considerations suggest either
hierarchical or quasi-degenerate neutrino masses.
While solar neutrino rates favour the small mixing angle MSW
oscillations (Wolfenstein 1978, Smirnov 1986), present data on the
recoil-electron spectrum prefer the large mixing solutions.
When interpreted in terms of neutrino oscillations, the observed
atmospheric neutrino zenith-angle-dependent deficit clearly indicates
that the mixing involved is maximal (Gonzalez-Garcia 2001).
Adding information from reactor experiments one concludes that the
third angle amongst the three neutrinos is small (Apollonio 1999).
Thus, altogether, we have the intriguing possibility that, unlike the
case of quarks, neutrino mixing is bi-maximal (Barger 1998, Davidson
1998, de~Gouvea 2000, Chankowski 2000fp, Hirsch 2000) which could be
tested at the upcoming long-baseline experiments or at a neutrino
factory experiment (Quigg 1999) or at the proposed KamLAND
experiment (De Braeckeleer 2000).

In addition to the above, there is also a long history of searches for
neutrino oscillations at accelerators.
Except for the unconfirmed hint provided by the LSND experiment
(Athanassopoulos 1998,~Smith 2000), accelerator searches have so--far
been negative.
The resulting limits, however, are not very restrictive on the scale
of the indications from underground experiments and I will not
discuss them any further.
Barring exotic neutrino conversion mechanisms the hint of the LSND
experiment together with the solar and atmospheric data require {\sl
  three mass scales}, hence the need for a fourth light neutrino,
which must be sterile~(Peltoniemi 1993, Caldwell 1993, Liu 1998,
Hirsch 2000, Giunti 2000). The most attractive possibility is to have,
out of the four neutrinos, two of them lie at the solar neutrino
scale, with the other two maximally-mixed neutrinos at the LSND
scale~(Peltoniemi 1993, Caldwell 1993, Hirsch 2000).  These schemes
have distinct implications at future solar \& atmospheric neutrino
experiments with sensitivity to neutral current neutrino interactions
such as SNO.  Cosmology can also place restrictions on these
four-neutrino schemes~(Raffelt 1999).

\section{Indications for New Physics}
\vskip .1cm

The most solid hints in favour of new physics in the neutrino sector
come from underground experiments on solar (Suzuki 2000) and
atmospheric (Sobel 2000, Becker-Szendy 1992) neutrinos.  The most
recent SK data correspond to 1117--day solar and 1144--day (71
kton-yr) atmospheric data samples, respectively. There are also new
data from Soudan-2 (5.1 kton-yr) and MACRO.

\subsection{Solar Neutrinos}
\vskip .1cm
 
\begin{figure}[t]
\centerline{\protect\hbox{
\psfig{file=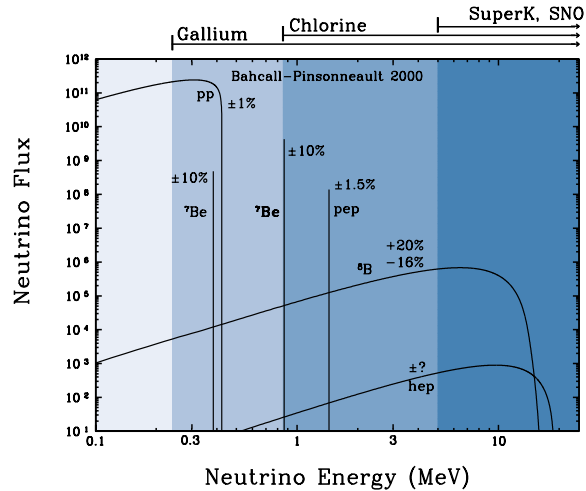,height=9cm,width=7cm,angle=-90}}}
\caption{Bahcall--Pinsonneault solar neutrino fluxes}
\label{solflux}
\end{figure}

\begin{figure}[t]
\centerline{\protect\hbox{
\psfig{file=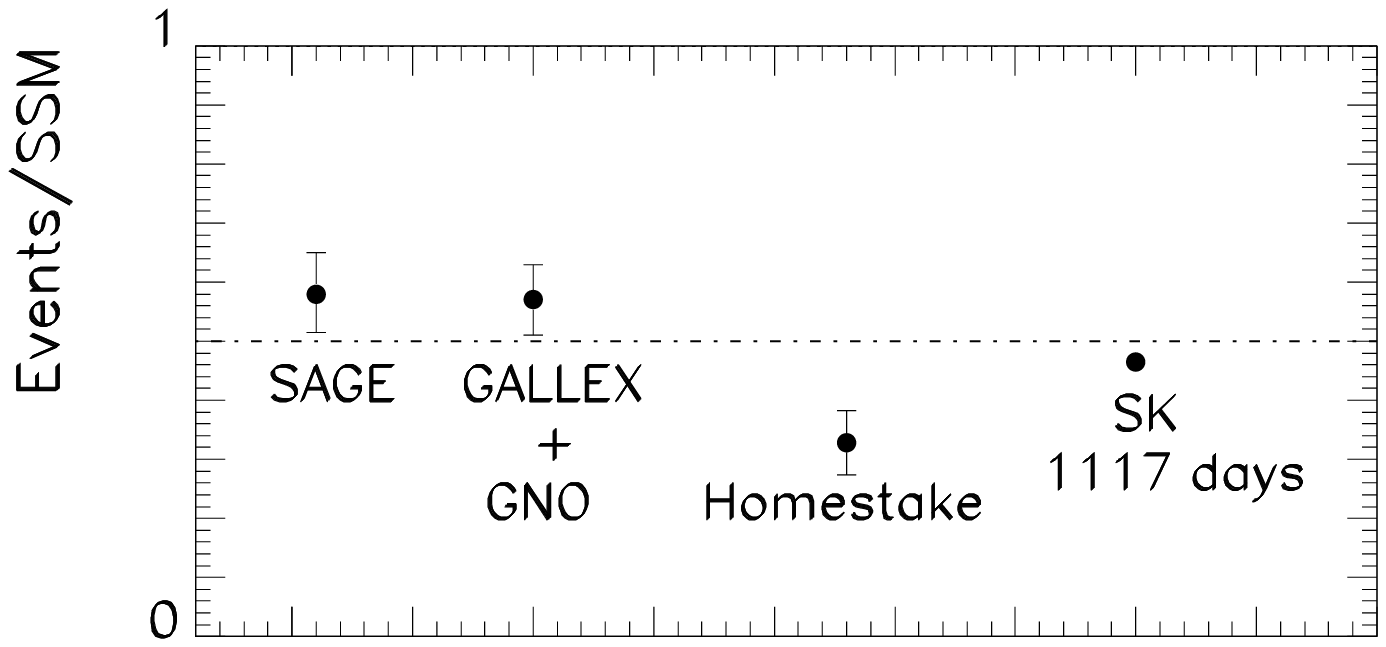,height=6cm,width=9cm}}}
\caption{Solar neutrino event rates normalized to SSM prediction.}
\label{solrates}
\end{figure}

\begin{figure}[t]
\centerline{\protect\hbox{
\psfig{file=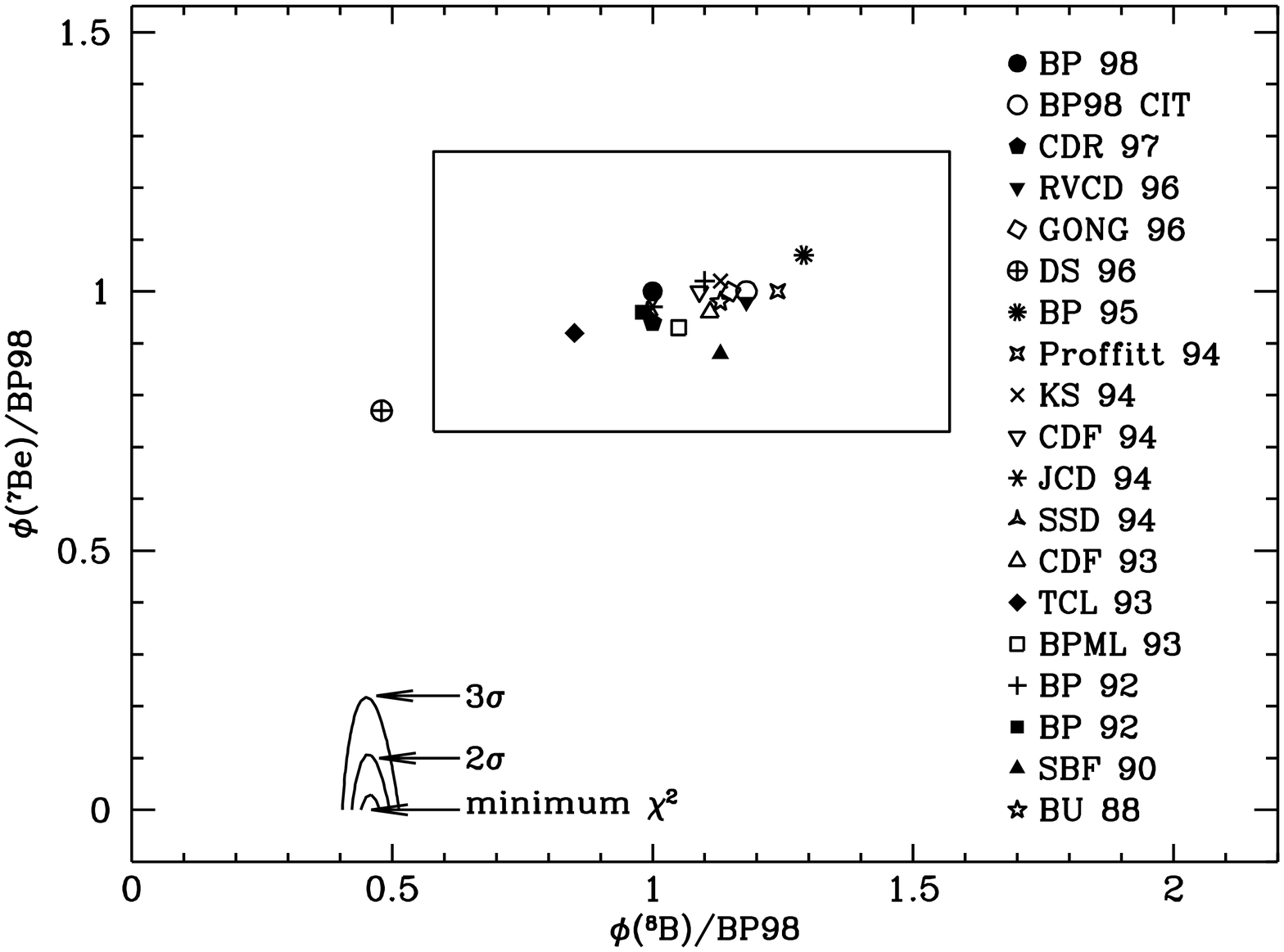,height=6cm,width=9cm}}}
\caption{$^7$Be and $^8$B neutrino fluxes: theory  versus  data }
\label{78}
\end{figure}

Our sun produces \neq's through various nuclear reactions which take
place in its interior. The predicted spectrum of solar neutrinos is
illustrated in \fig{solflux}, taken from~(Bahcall 1998).  I will
refer to this model as ``the'' SSM.
Solar neutrinos are detected either with geochemical methods (the
$\nu_e + ^{37}Cl \to ^{37}Ar + e^-$ reaction at the Homestake
experiment and the $\nu_e + ^{71}Ga \to ^{71}Ge + e^-$ reaction at the
Gallex, Sage and GNO experiments) or through $\nu_e\,e^-$ scattering on
water, using Cerenkov techniques at Kamiokande and Super-Kamiokande.
As summarized in \fig{solrates} all experiments observe a deficit of
30 to 60 \% whose energy dependence follows mainly from the lower
Chlorine rate.  Note that \fig{solrates} includes the latest results
from SK, SAGE \& GNO presented at $\nu$2000, but not the first results
from SNO.
It is convenient to present the predictions of various standard solar
models in terms of the $^7$Be and $^8$B neutrino fluxes, normalized to
the SSM predictions~(Bahcall 1998), as seen in \fig{78}, which
includes most of the existing solar models.

On the other hand the values of the fluxes indicated by measured
neutrino event rates are shown by the contours in the lower-left part
of the figure, with a negative best-fit $^7$Be neutrino flux!  This
discrepancy strongly suggests the need for new particle
physics~(Bahcall 1994).  Since possible non-standard astrophysical
solutions are rather constrained by helioseismology studies (Bahcall
1998) one is led to assume the existence of neutrino conversions, such
as those induced by very small neutrino masses.

The high statistics of SK after 1117 days of data--taking also
provides very useful information on the recoil electron energy
spectrum with event rates given for 18 bins starting at 5.5
MeV~\footnote{They have also reported results of a lower energy bin 5
  MeV $<E_e<$5.5 MeV, but due to systematic errors this is not yet
  included in the analysis.}.
\begin{figure}
\vglue -.5cm
\centerline{\protect\hbox{
\psfig{file=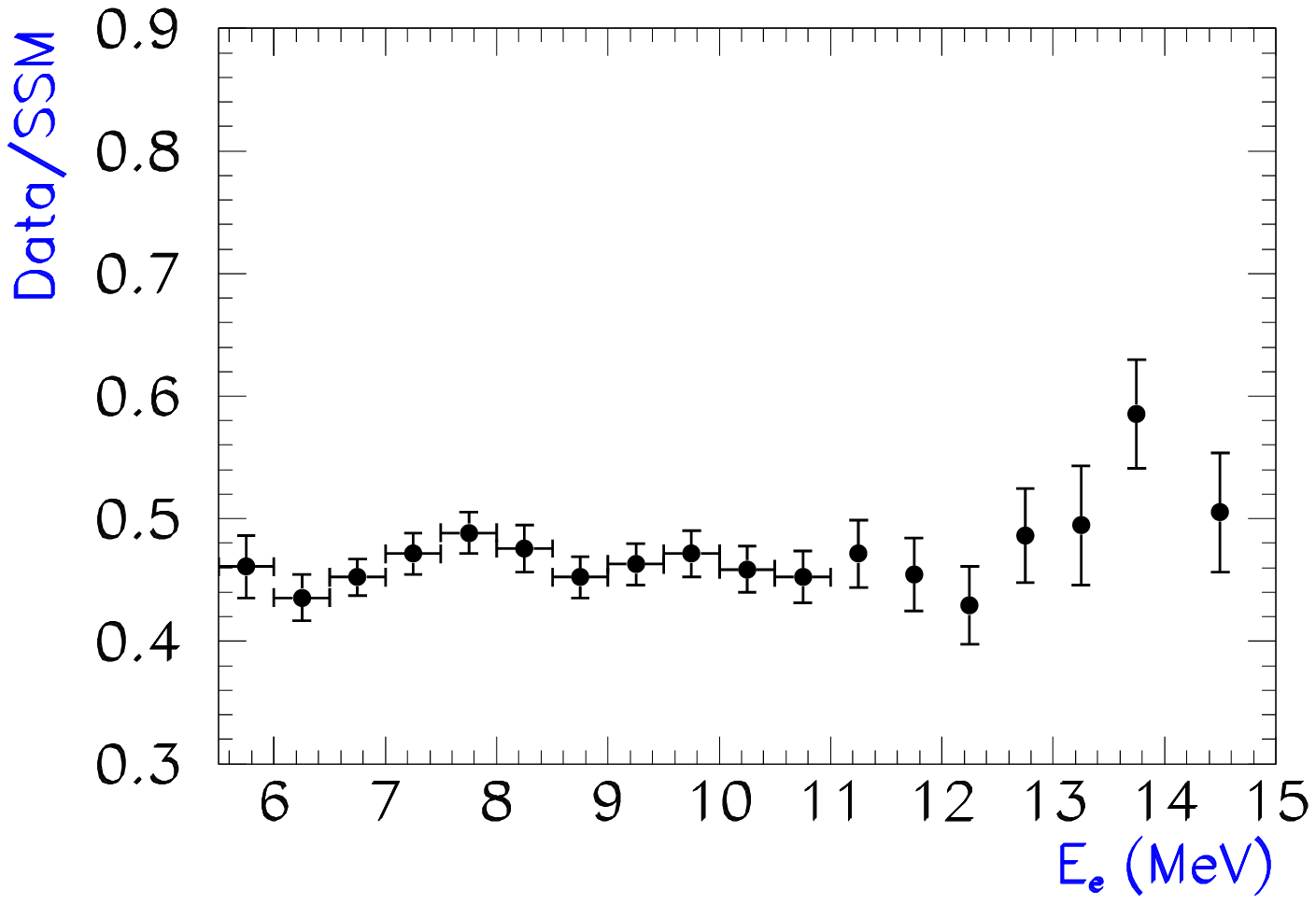,height=6cm,width=9.5cm}}}
\caption{Measured recoil electron energy spectra~(Suzuki 2000).  }
\label{spec1117}
\end{figure}
The spectrum in \fig{spec1117} is well described by the flat
hypothesis $\chi^2_{flat}=13/(17\rm{dof})$, in contrast with hints
from previous 825--day sample.
Moreover, SK measures the zenith angle distribution (day/night effect)
which is sensitive to the effect of the Earth matter in the neutrino
propagation.  
\begin{figure}[htbp]
\vglue -.2cm
\centerline{\protect\hbox{
\psfig{file=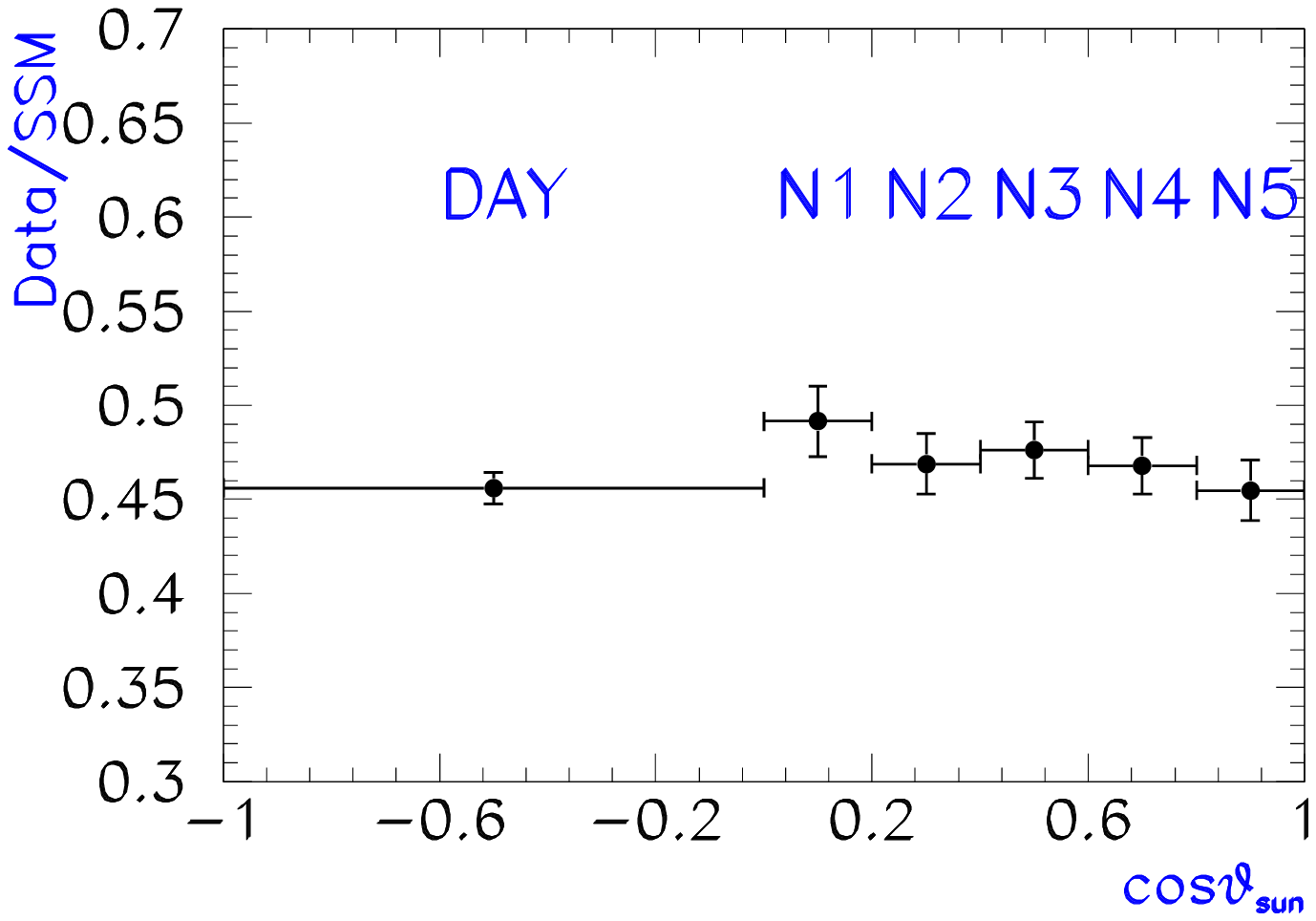,height=6cm,width=9cm}}}
\caption{Zenith angle distribution normalized to the SSM prediction.}
\label{}
\end{figure} 
One sees a slight excess of events at night, but the corresponding day
versus night asymmetry $A_{D/N}=\frac{D-N}{\frac{D+N}{2}}=-0.034\pm
0.022 \pm 0.013$ is only 1.3$\sigma$ away from zero.
In order to combine this day--night information with the spectral
data, SK has also presented separately the measured recoil energy
spectrum during the day and during the night. This will be referred in
the following as the day--night spectra data and contains $2\times
18$ data bins.

Note that the absence of clear hints of spectral distortion, day-night
or seasonal variation implies that, {\sl per se}, they do not give any
clear indication for physics beyond the standard model.  From this
point of view, despite the increasing weight of such rate-independent
observables, the solar neutrino problem rests heavily on the rate
discrepancy.
Nevertheless, as we will see, rate-independent observables are already
playing an important r{\^o}le in selecting amongst different solutions
of the solar neutrino problem.

\subsection{Atmospheric Neutrinos}
\vskip .1cm

\begin{figure}[t]
\centerline{\protect\hbox{
\psfig{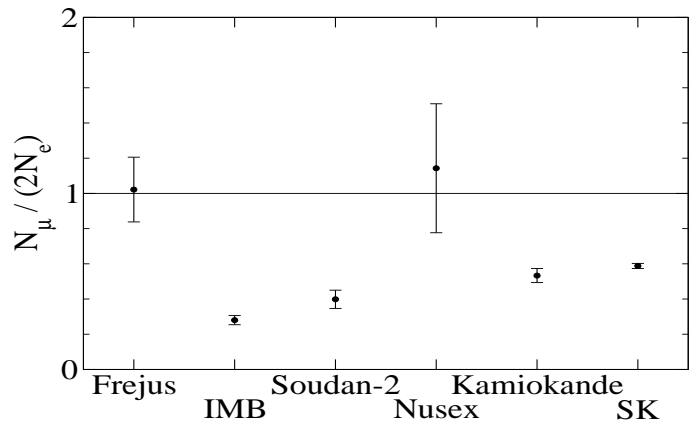}}}
\caption{Atmospheric neutrino event rates normalized 
to theory}
\label{atmrates}
\end{figure}

\begin{figure} 
\begin{center}
\includegraphics[width=10.5cm,height=8cm]{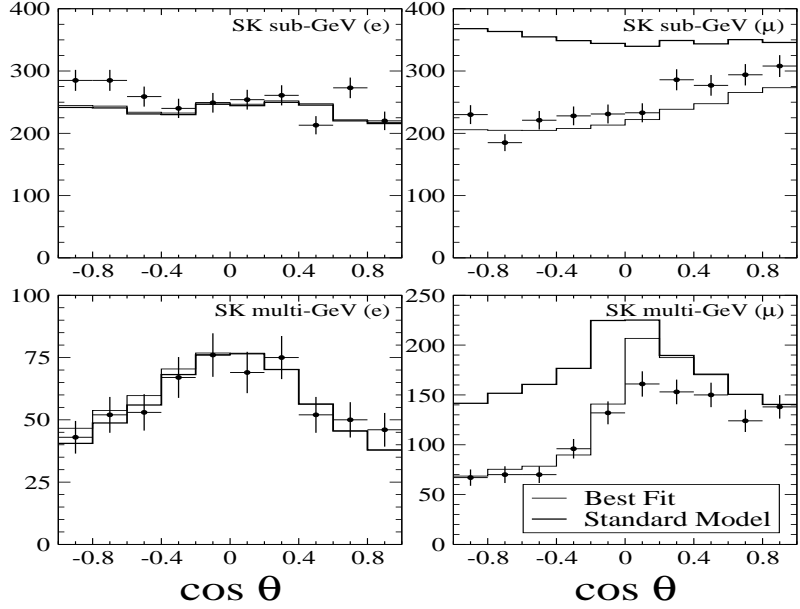}
      \end{center}
\caption{
  SK zenith-angle distributions for contained events versus
  theoretical expectations in the SM and within the oscillation
  hypothesis}
\label{fig:angcont}
\end{figure}

Neutrinos produced as decay products in hadronic showers from cosmic
ray collisions with nuclei in the upper atmosphere have been observed
in several experiments (Sobel 2000, Becker-Szendy 1992).
Although individual $\nu_\mu$ or $\nu_e$ fluxes are only known to
within $30\%$ accuracy, their ratio is predicted to $5\%$ over
energies that vary from 0.1~GeV to 100~GeV~(Gaisser 1998).
The long-standing discrepancy between the predicted and measured
$\mu/e$ ratio of the muon ($\nu_\mu + \bar{\nu}_\mu$) over the
electron atmospheric neutrino flux ($\nu_e+\bar{\nu}_e$)~(Gaisser
1995) found both in water Cerenkov experiments (Kamiokande, SK and
IMB) as well as in the iron calorimeter Soudan2 experiment is
illustrated in \fig{atmrates}
This evidence has now been strengthened by the fact that it exhibits a
strong {\sl zenith-angle dependence}~(Sobel 2000) as can be seen from
\fig{fig:angcont}.
The zenith-angle distributions for the Super-Kamiokande $e$--like are
shown in the left panels, both in the sub-GeV (upper panels) and
multi-GeV (lower panels) energy range.  The thick solid line is the
expected distribution in the SM.  It is consistent with the SM
expectations.
In contrast $\mu$--like events displayed in the right-panels show a
clear deficit of neutrinos coming from below, which is very suggestive
indeed of \nm oscillations.
In \fig{fig:angcont} we also give the predicted best-fit distributions
obtained in a global 3-neutrino oscillation description of the
data~(Gonzalez-Garcia 2001).
The thin full line is the prediction for the overall best-fit point of
the contained plus up-going atmospheric data sample, with
$\tan^2\theta_{13}=0.025$, $\Delta m^2_{32}=3.3\times 10^{-3}$ eV$^2$
and $\tan^2\theta_{23}=1.6$.  

\begin{figure}
\begin{center}
\mbox{\epsfig{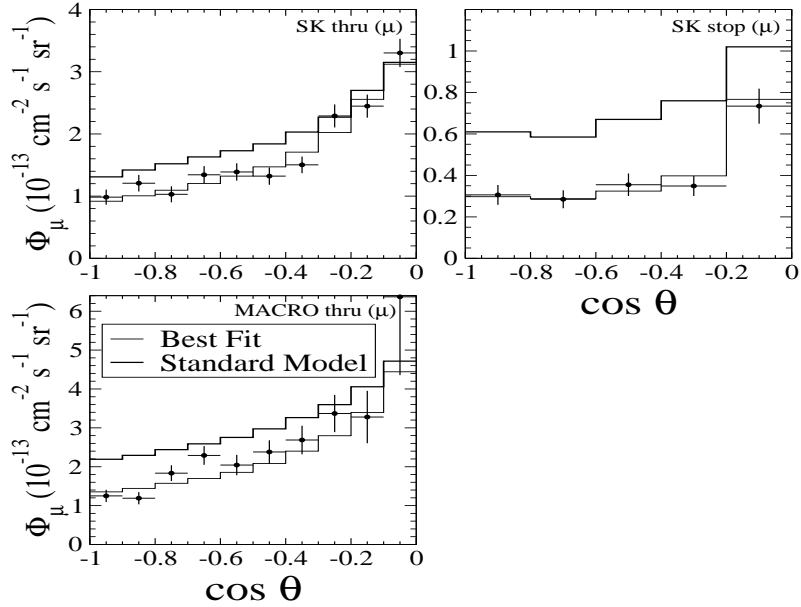}} 
\end{center}
    \caption{
      Zenith-angle distributions for upward-going muon events in
      Super-Kamiokande and MACRO.}
    \label{fig:angup}
\end{figure}

Zenith-angle distributions have also been recorded for upward-going
muon events in Super-Kamiokande and MACRO, as illustrated in
\fig{fig:angup}.
The thick solid line is the expected distribution in the SM, while the
thin full line corresponds, as in \fig{fig:angcont}, to the prediction
for the overall global best-fit point.

\section{Three--neutrino fits}
\vskip .1cm

The most economical joint description of solar and atmospheric
anomalies involves oscillations amongst all three known types of
neutrinos.
Here I summarize the results of the recent global analysis
(Gonzalez-Garcia 2001) of the solar~(Suzuki 2000), atmospheric~(Sobel
2000, Becker-Szendy 1992) and reactor~(Apollonio 1999) neutrino data
in terms of three--neutrino oscillations.
The present discussion goes beyond previous three--neutrino
oscillation analyses including only solar~(Fogli 2000) or only
atmospheric neutrino data~(Fogli 1999).
It also updates some joint studies~(Barger 1980) which do not take
into account the latest SK data. As we saw the most recent solar
neutrino rates include the 1117 day SK data sample on the recoil
electron energy spectra for day and night periods.
On the other hand the atmospheric data sample includes not only
contained events but also the upward-going $\nu$-induced muon fluxes.
In addition to previous Frejus, IMB, Nusex, and Kamioka data we use
the most recent 71 kton-yr (1144 days) SK data set, the 5.1 kton-yr
contained events of Soudan2, and the results on up--going muons from
the MACRO detector.

The pattern of neutrino oscillations expected in any fundamental
(gauge) theory of neutrino mass is determined by the structure of the
lepton mixing matrix~(Valle 1991, Bilenkii 1999).
For the simplest three--neutrino theories this is in general
characterized by three mixing angles and three CP violating
phases (Schechter 1980a).
The latter include, in addition to the Dirac-type phase analogous to
that of the quark sector, two extra physical~(Schechter 1981a) phases
associated to the Majorana character of neutrinos.  Conservation of CP
implies that Dirac phases are zero modulo $\pi$, while Majorana phases
are zero modulo$\pi/2$ (Wolfenstein 1981, Schechter 1981b).
For our following discussion all three phases are set to zero.  In
this case the mixing matrix can be conveniently chosen in the form
(Schechter 1980a)
\begin{equation}
\left(
    \begin{array}{ccc}                                
        c_{13} c_{12}                
        &  s_{12} c_{13} 
        &  s_{13} \\
        -s_{12} c_{23} - s_{23} s_{13} c_{12} 
        &  c_{23} c_{12} - s_{23} s_{13} s_{12}
        &  s_{23} c_{13} \\
         s_{23} s_{12} - s_{13} c_{23} c_{12}
        &  -s_{23} c_{12} - s_{13} s_{12} c_{23}
        &  c_{23} c_{13} 
    \end{array}\right)
\label{eq:evol.2} 
\end{equation}

The joint study of solar and atmospheric neutrino oscillations is
characterized by a five-dimensional parameter space
\begin{equation} \begin{array}{ll} 
   \label{oscpardef}
    \Delta m^2_{\odot} & \equiv \Delta m^2_{21} = m^2_2 - m^2_1 \\
    \Delta m^2_{atm}   & \equiv \Delta m^2_{32} = m^2_3 - m^2_2 \\
    \theta_{\odot}     & \equiv \theta_{12} \\
    \theta_{atm}       & \equiv \theta_{23} \\
    \theta_{reactor}   & \equiv \theta_{13}
\end{array} 
\end{equation}
where all mixing angles are assumed to lie in the full range from
$[0,\pi/2]$. 

From the required hierarchy in the splittings $\Delta m^2_{atm} \gg
\Delta m^2_{\odot}$ indicated by the solutions to the solar and
atmospheric neutrino anomalies (see below) it follows that the
analyses of solar data constrain three of the five independent
oscillation parameters, namely, $\Delta m^2_{21}, \theta_{12}$ and
$\theta_{13}$ since for most cases oscillations over the atmospheric
scale  average out.
Conversely, from the point of view of the atmospheric data analysis
one can effectively assume that the lighter neutrinos become
degenerate so that one can rotate away the corresponding angle
$\theta_{12}$. The leptonic mixing matrix takes on the simplified
form~(Schechter 1980b)
\begin{equation}
    {\bf R} =\left(
    \begin{array}{ccc}                                 
        c_{13}          & 0       & s_{13} \\
        - s_{23} s_{13} & c_{23}  & s_{23} c_{13} \\
        - s_{13} c_{23} & -s_{23} & c_{23} c_{13}
    \end{array}\right); 
\label{eq:evolap.1}
\end{equation}
As a result only three oscillation parameters: $\Delta m^2_{32}$,
$\theta_{23}$ and $\theta_{13}$ are necessary to describe the
3-neutrino propagation of atmospheric neutrinos.

It follows from the above discussion that $\theta_{13}$ is the only
parameter common to both analyses and may potentially allow for some
``cross-talk'' between the two sectors.
It is known that for $\Delta m^2_{32} \gg \Delta m^2_{12}$ and
$\theta_{13}=0$ the atmospheric and solar neutrino oscillations
decouple in two 2--$\nu$ oscillation scenarios. In this respect our
results also contain as limiting cases the pure two--neutrino
oscillation scenarios and update previous analyses on atmospheric
neutrinos~(Fornengo 2000, Foot 1998) and solar
neutrinos~(Gonzalez-Garcia 2000a, Bahcall 1998).
 
In order to compute the solar neutrinos survival probabilities for any
value of the neutrino mass and mixing the full expression for the
survival probability has been used, without appealing to the usual
approximations whose validity defines the MSW~(Smirnov 1986) or the ``
just-so''(Glashow 1987) regime.
The treatment of neutrino oscillations is therefore unified, with MSW
and vacuum oscillations considered on the same footing.  Likewise, we
include in our description conversions with $\theta_{12} >
\pi/4$~(Gonzalez-Garcia 2000b).
Results are found by numerically solving the Schrodinger neutrino
evolution equation in the Sun and the Earth matter, using the electron
number density of BP2000 model~(Bahcall 2000) and the Earth density
profile given in the Preliminary Reference Earth Model
(PREM)~(Dziewonski, 1981).\\

{\bf Reactor limits}\\

\begin{figure} 
\begin{center} 
\mbox{\epsfig{file=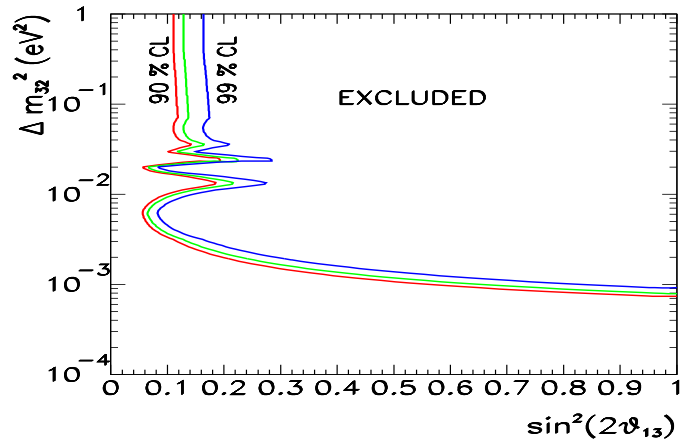,height=6cm,width=9cm}}
\end{center} 
\caption{Excluded region in  $\Delta{m}^2_{32}$ and $\sin^2(2\theta_{13})$  
from the non observation of oscillations by the CHOOZ reactor.} 
\label{fig:chooz} 
\end{figure}

To start the summary of the global 3-neutrino fits we first note that,
of all laboratory searches for neutrino oscillation, reactors provide
the most sensitive one when it comes to comparing with the strong
hints from underground experiments. Thus we will first consider the
region of oscillation parameters which can be excluded from reactor
experiments.  The restrictions on $\Delta{m}^2_{32}$ and
$\sin^2(2\theta_{13})$ that follow from the non observation of
oscillations at the CHOOZ reactor experiment are shown in
\fig{fig:chooz}, taken from~(Gonzalez-Garcia 2001).
The curves represent the 90, 95 and 99\% CL excluded region defined
with 2 d.~o.~f. for comparison with the CHOOZ published results.
In what follows we will compare this direct bound with those obtained
from a global analysis of solar and atmospheric data, as well as
consider its effect in combination with the latter.\\

{\bf Solar data fit}\\

\begin{figure}
 \vglue -.5cm 
\begin{center}
\includegraphics[width=12cm,height=7cm]{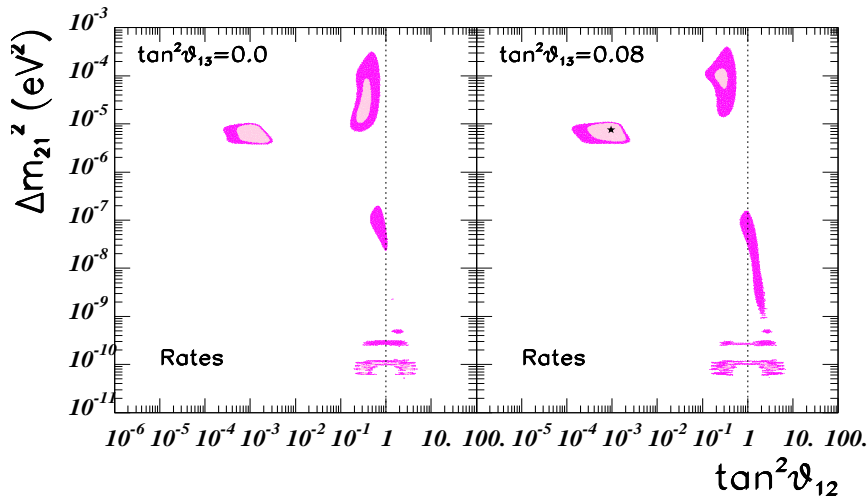}
      \end{center}
    \vspace{3mm}
\caption{3-$\nu$ oscillation
  regions \texttt{allowed} at 99\% (darker) and 90\% CL (lighter) by
  the latest measurements of the total solar neutrino event rates.
  The best--fit point is denoted as a star. }
\label{fig:3sol_r}
\end{figure}

We first present the allowed regions of solar oscillation parameters
$\theta_{12}$, $\Delta m^2_{21}$ as a function of $\theta_{13}$.  All
plots are taken from~(Gonzalez-Garcia 2001) where all details can be
found.
In \fig{fig:3sol_r} we give the allowed three--neutrino oscillation
regions in $\Delta{m}^2_{21}$ and $\tan^2\theta_{12}$ from the
measurements of the total event rates at the Chlorine, Gallium,
Kamiokande and Super-Kamiokande experiments.  The different panels
represent the \texttt{allowed} regions at 99\% (darker) and 90\% CL
(lighter) obtained as sections for fixed values of the mixing angle
$\tan^2\theta_{13}$ of the three--dimensional volume defined by
$\chi^2-\chi^2_{min}$=6.25 (90\%), 11.36 (99\%).
The best--fit point is denoted as a star. It occurs, as expected, for
the small mixing MSW solution (SMA) simply because this is the
situation which most strongly suppresses the unwanted $^7$Be neutrino
flux. It is characterized by a non-zero $\theta_{13}$ value. For
higher $\theta_{13}$, the description worsens. Although relatively
weak, the limit on $\theta_{13}$ from solar data is totally
independent on the allowed range of the atmospheric mass difference
$\Delta m^2_{32}$.

\begin{figure}
\vglue -.5cm 
\begin{center}
\includegraphics[width=12cm,height=7cm]{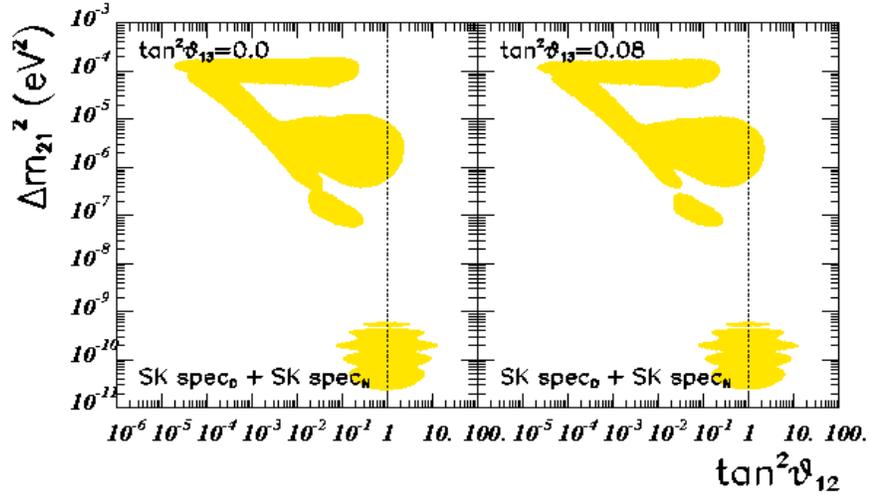}
      \end{center}
    \vspace{3mm}
\caption{3-$\nu$ oscillation solar oscillation regions 
  \texttt{excluded} at 99\% CL by the day--night spectra measurement. }
\label{fig:3sol_ex}
\end{figure}

The three--neutrino solar oscillation regions \texttt{excluded} by the
measurement of the day--night spectra data in the Super-Kamiokande
1117-day data sample is illustrated in \fig{fig:3sol_ex}.

\begin{figure}
\vglue -.5cm 
\begin{center}
\includegraphics[width=12cm,height=7cm]{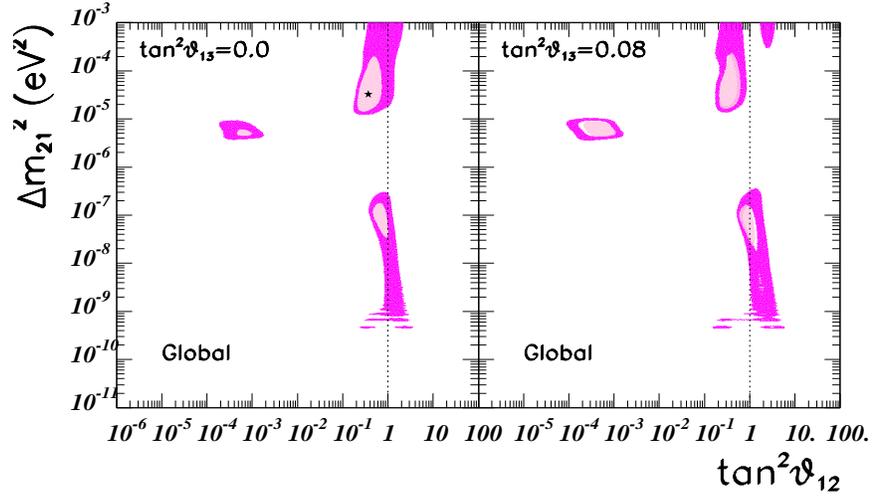}
      \end{center}
    \vspace{3mm}
\caption{3-$\nu$ oscillation
  regions \texttt{allowed} by all of the solar neutrino data. The
  best--fit point is denoted as a star.}
\label{fig:3sol_rspdn}
\end{figure}

Finally, the \texttt{allowed} three--neutrino solar oscillation
regions in $\Delta{m}^2_{21}$ and $\tan^2\theta_{12}$ which follows
from the global analysis of solar neutrino data is presented in
\fig{fig:3sol_rspdn}. The best--fit point is denoted as a star.

\begin{figure}
\vglue -.3cm 
\begin{center} 
\includegraphics[width=12cm,height=7cm]{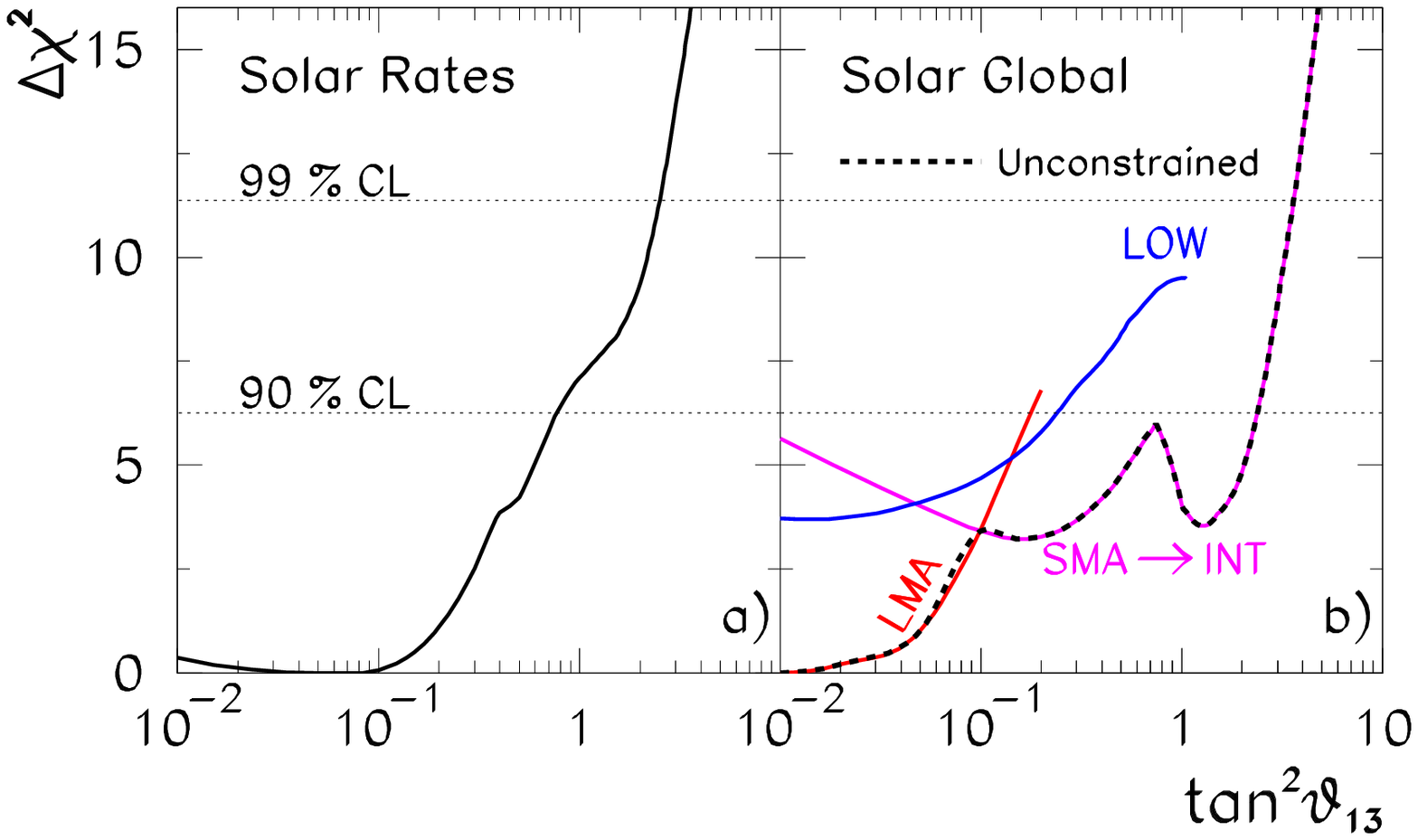} 
\end{center} 
\caption{Relative quality of 
three--neutrino solutions to the solar neutrino problem. }
\label{fig:chisol} 
\end{figure}

The relative quality of the various oscillation solutions of the solar
neutrino problem is illustrated in \fig{fig:chisol}.
This figure gives $\Delta \chi^2$ as a function of $\tan^2\theta_{13}$
from the 3--neutrino analysis of the solar data.  The dotted
horizontal lines correspond to the 90\%, 99\% CL limits. 
The left panel corresponds the analysis of total rates only, while the
right panel corresponds to the global analysis.  The dotted horizontal
lines correspond to the 90\%, 99\% CL limits.
Though all the various SMA, LMA, LOW and vacuum solutions are still
acceptable global descriptions of the solar data, they are not equally
good.
From the right panel in \fig{fig:chisol} we conclude that, for small
enough $\tan^2\theta_{13}$ values (so that we gets effectively a
two-neutrino scheme) the best solution is the LMA solution is the
best, while SMA is the worst.
One notices also that rate-independent observables, such as the
electron recoil energy day-night spectra, are playing an increasing
role in discriminating between different solutions to the solar
neutrino problem, pushing the best--fit point towards the LMA
solution~(Gonzalez-Garcia 2001), a trend already noted in earlier
2-neutrino analyses due to the same reason~(Gonzalez-Garcia 2000a).

Another issue which could play a more significant role in future
investigations is that of seasonal variations, expected in the just-so
regime and, more subtly, also in the MSW large mixing solutions, LMA
and LOW.
The first would result from the eccentricity of the Earth's orbit
around the Sun and could be tested by searching for seasonal
variations in the $^7$Be neutrino flux at the Borexino experiment, and
possibly at KamLAND~(de~Gouvea 1999).
The latter would result from the regeneration effect at the Earth
(day-night effect) and might be tested through time variations of
event rates at GNO and Borexino~(de~Holanda 1999).

An interesting theoretical issue is the possible effect of random
fluctuations in the solar matter density~(Balantekin 1996, Nunokawa
1996, Bamert 1998) on the solar neutrino event rates.
The existence of such noise fluctuations at a few percent level is not
excluded by present helioseismology studies.
The correlation length $L_0$ associated with the scale of the
fluctuation can be assumed to lie between the mean free path of the
electrons in the solar medium, $l_{free} \sim 10 $ cm, and the
neutrino oscillation length in matter, $\lambda_m$, e.~g.~ $ l_{free}
\ll L_0 \ll \lambda_m$.
Even small fluctuations can have an important effect on averaged solar
neutrino survival probabilities, especially for small solar mixing
angles~~(Balantekin 1996, Nunokawa 1996).
The fluctuations can affect the $^7$Be neutrino component of the solar
neutrino spectrum, implying that Borexino can probe, at some level of
precision, the magnitude of solar matter density fluctuations, thus an
additional motivation for the experiment (Arpesella 1991).\\

{\bf Atmospheric and reactor data fit}\\

\begin{figure} 
\begin{center}
\includegraphics[width=12cm,height=6.5cm]{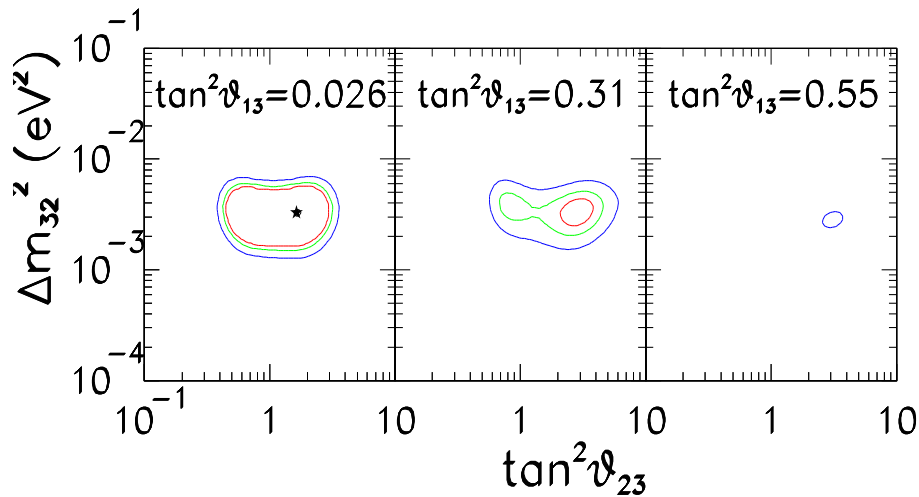}
      \end{center}
    \vspace{3mm}
\caption{90, 95 and 99\% CL regions in $(\tan^2\theta_{23}, \Delta m^2_{32})$  
  \texttt{allowed} by the combination of all atmospheric data.  }
\label{fig:sk}
\end{figure}

\begin{figure} 
\begin{center}
\includegraphics[width=12cm,height=6.5cm]{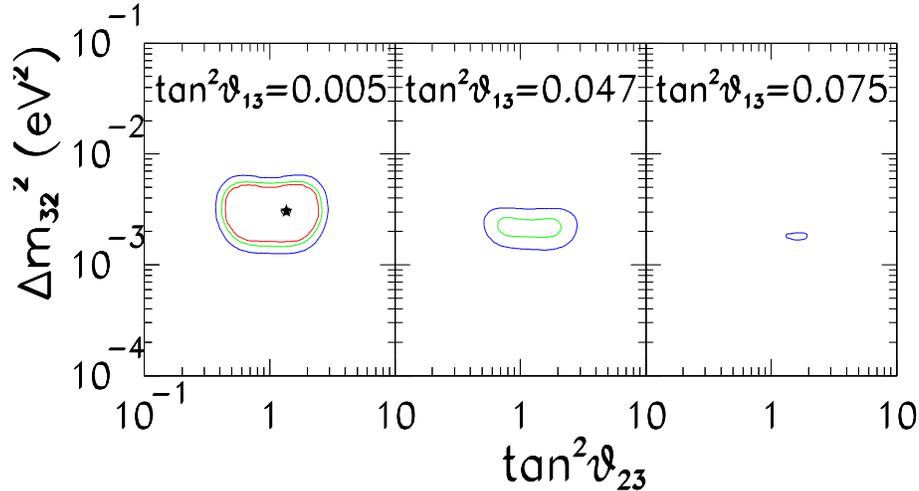}
      \end{center}
    \vspace{3mm}
\caption{ 90, 95 and 99\% CL
  3--$\nu$ regions in $(\tan^2\theta_{23}, ~\Delta m^2_{32})$ allowed
  by the combination of all neutrino data.  }
\label{fig:all}
\end{figure}

Here I present the allowed regions of atmospheric oscillation
parameters $\theta_{23}$, $\Delta m^2_{32}$ for different values of
$\theta_{13}$, common to solar and atmospheric analyses.  All plots
are taken from (Gonzalez-Garcia 2001) where details can be found.
In \fig{fig:sk} we display the \texttt{allowed} $(\tan^2\theta_{23},
\Delta m^2_{32})$ regions for different $\tan^2\theta_{13}$ values,
that follow from the combination of Super-Kamiokande atmospheric
neutrino events.  The regions refer to 90, 95 and 99\% CL.  The
best--fit point is denoted as a star and corresponds to
$\tan^2\theta_{13} = 0.026$.
In \fig{fig:all} we present the three--neutrino regions in
$(\tan^2\theta_{23}, ~\Delta m^2_{32})$ \texttt{allowed} by the
combination of all atmospheric neutrino data plus Chooz.  The
best--fit point is denoted as a star.
Comparing \fig{fig:sk} and \fig{fig:all} one can see the weight of the
reactor neutrino bound on the global analysis. The best global fit has
$\tan^2\theta_{13} = 0.005$ and for $\tan^2\theta_{13} = 0.075$ even
the 99\% CL allowed disappears.
Notice the important complementarity between atmospheric data and the
reactor limits on $\theta_{13}$ since the latter apply only for  
$\Delta m^2_{32} \gsim 10^{-3} $ eV$^2$.
\newpage
{\bf Combining solar, atmospheric and reactor data}\\

\begin{figure} 
\begin{center}
\includegraphics[width=12cm,height=7cm]{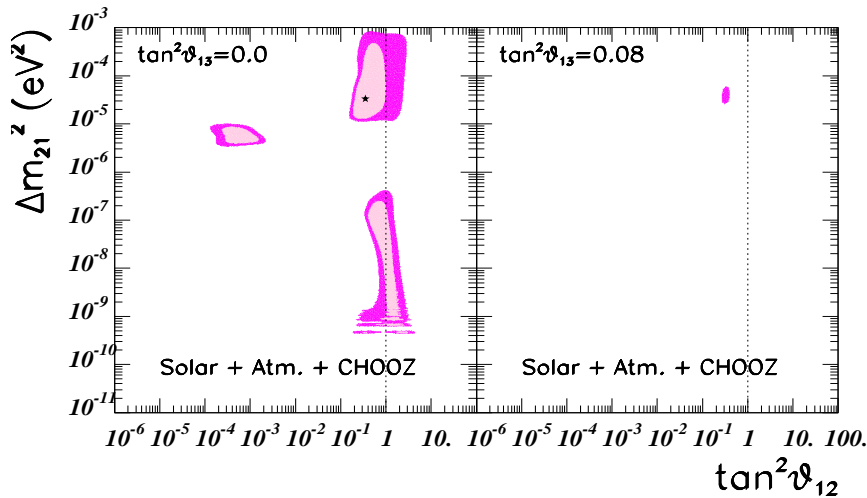}
      \end{center}
    \vspace{3mm}
\caption{($\Delta{m}^2_{21}$, $\tan^2\theta_{12}$) regions   allowed 
  by the global solar, atmospheric and reactor neutrino
  data analysis. }
\label{fig:glosol}
\end{figure}

We also obtain the allowed ranges of parameters from the full
five--dimensional combined analysis of all of the above neutrino data.
In \fig{fig:glosol} we give the regions in $\Delta{m}^2_{21}$ and
$\tan^2\theta_{12}$ \texttt{allowed} by the global analysis of solar,
atmospheric and reactor neutrino data. The left panel gives the
regions for the unconstrained analysis defined in terms of the
increases of $\Delta\chi^2$ for 5~d.o.f.\ from the global best fit
point denoted as a star. The right panel shows the values of
$\tan^2\theta_{13}$ beyond which the 99\% CL region starts to
disappear.

\section{Putting the pieces together }
\vskip .1cm
 
Together with the solar neutrino data, the angle-dependent atmospheric
neutrino deficits provide a strong evidence for physics beyond the
Standard Model. 
Small neutrino masses provide the simplest and most generic
explanation of the data.
Theoretical neutrino mass models fall into two classes, which I call
bottom-up and top-bottom. They can lead to either hierarchical
~(de~Gouvea 2000, Hirsch 2000) or quasi-degenerate~(Chankowski 2000)
neutrinos.
Top-bottom approaches inspired on the idea of Unification typically
employ either a see-saw mechanism, or high dimension operators. As
examples of minimalistic models I mention two we have recently
investigated, both of which require a large mixing solution to the
solar neutrino problem~(de~Gouvea 2000, Chankowski 2000).

Supersymmetry with bilinear breaking of R--parity provides a simple
bottom-up-type model which allows for the possibility of probing the
neutrino mixing, as indicated by the underground experiments, within
the context of high--energy collider experiments such as the LHC
~(Hirsch 2000).
For additional models and for models involving specific Yukawa
textures see~(Davidson 1998, Lola 1998, Altarelli 1998).

Last, but not least, I mention that it is impossible to reconcile
solar, atmospheric and reactor data with the LSND hint in terms of
neutrino oscillations without a light sterile neutrino~(Peltoniemi
1993, Caldwell 1993) and in ref.~(Hirsch 2000) we give an updated
discussion in the context of an interesting model.
For more references see~(Liu 1998) and for a more complete
discussion of limits on 4-neutrino models see~(Giunti 2000).
If light sterile neutrinos exist they should be probed at
neutral-current-sensitive solar \& atmospheric neutrino experiments
such as SNO~(Gonzalez-Garcia 2000c).

\section{Conclusion}
\vskip .1cm

Within the neutrino oscillation framework present solar and
atmospheric data suggest the intriguing possibility of bi-maximal
neutrino mixing, which explicitly illustrates the sharp contrast
between the lepton and quark sectors of the theory. 
With good luck this could be checked on the one hand at the upcoming
long-baseline experiments or at a neutrino factory~(Quigg 1999) and,
on the other, via the search of seasonal effects in solar neutrinos,
e.~g.~at the proposed KamLAND experiment~(De Braeckeleer 2000).
In certain models~(Hirsch 2000) one may test the neutrino mixing
angles involved in the explanation of the neutrino anomalies at
high--energy collider experiments~(Porod 2000), illustrating an
amusing synergy between accelerator and underground experiments.
On the other hand if the LSND result stands the test of time, this
would be a strong indication for the existence of a light sterile
neutrino which would be another radical difference between leptons and
quarks.
Let me mention however that the neutrino oscillation interpretation of
solar neutrino anomalies is still far from unique.
For example, resonant spin flavor precession~(Akhmedov 1988) induced
by transition magnetic moments~(Schechter 1981b). For a recent fit of
solar data see ref.~(Miranda 2001).
From this point of view, it is still too early for a precision
determination of neutrino properties from underground experiments.
Last, but not least, let me mention that in fifty years of weak
interaction physics an answer to the most fundamental question about
the nature - Dirac versus Majorana - of neutrinos has so far defied
all experimental attempts. 

\vskip .5cm

\noindent{\bf Acknowledgements:}  \\

This work was supported by Spanish DGICYT under grant PB98-0693, by
the European Commission TMR network HPRN-CT-2000-00148 and by the
European Science Foundation network grant N.  86.

\end{document}